\documentclass[anz]{cupaus}

\usepackage{bm}
\usepackage{amsmath}
\usepackage{hyperref}
\usepackage{graphicx}
\usepackage[numbers]{natbib}

\renewcommand{\doi}[1]{\href{https://doi.org/#1}{doi:#1}}

\begin{document}

\verso{Critical Triplet of the Orr--Sommerfeld Operator}
\recto{Critical Triplet of the Orr--Sommerfeld Operator}

\title{The Quad--Precision Critical Triplet of the Orr--Sommerfeld Operator}

\cauthormark 
\author[1]{Stephen~J.~Walters}

\address[1]{Mathematics Department, University of Tasmania, 
Tasmania~7005, \textsc{Australia}.
\email{stephen.walters@utas.edu.au}
}    

\pages{1}{13}

\begin{abstract}
Plane Poiseuille flow serves as a fundamental benchmark for hydrodynamic stability solvers and code verification. However, high-Reynolds-number stability formulations face severe numerical bottlenecks, including extreme operator ill-conditioning, catastrophic cancellation in finite-difference Jacobians, and precision limits. This paper presents a robust, quadruple-precision computational framework utilizing two independent methods -- a global spectral Galerkin scheme with analytical eigenvalue derivatives and a local Riccati shooting method. By deriving exact, analytically evaluated Jacobians and residuals, the framework eliminates the significant numerical error of finite-difference approximations, achieving flawless quadratic convergence down to the quad-precision floor. Cross-verified to 28 digits, the resulting extended-precision triplet ($Re_c=5772.2\, 21816\, 20969\, ... $, $\alpha_c = 1.0205\, 47449\, 28533\, ...$, 
$c_r = 0.26400\, 02604\, 78798\, ...$) provides an authoritative numerical benchmark. Furthermore, the accompanying open-source, high-precision code architecture establishes a careful verification suite for modern Navier-Stokes as well as magnetohydrodynamic (MHD) codes in their zero-magnetic-field limit.
\end{abstract}

\keywords[2020 \textit{Mathematics subject classification}]{76E05}

\keywords[\textit{Keywords and phrases}]{Plane Poiseuille flow, Orr--Sommerfeld equation, Hydrodynamic stability, Chebyshev spectral methods}

\maketitle

\section{Introduction}
\label{sec:intro}

\subsection{Background and Governing Equations}
The study of plane Poiseuille flow (PPF) has been a central theme of hydrodynamic stability research since Osborne Reynolds' 1883 pioneering work on transition to turbulence \citep{reynolds1883}. The stability of this flow, governed by the Orr-Sommerfeld equation, derived independently by Orr \cite{orr1907} in 1907 and Sommerfeld \citep{sommerfeld1908} in 1908, is one of the most fundamental and extensively studied paradigms in fluid mechanics. Due to its singular nature at high Reynolds numbers ($Re$), it historically served as the primary proving ground for asymptotic stability theory and early computational fluid dynamics. The Orr-Sommerfeld equation is a deceptively simple linear fourth-order eigenvalue problem that hides extreme mathematical complexities. Because the viscous terms are multiplied by $1/Re$, the equation becomes singularly perturbed and increasingly stiff at high Reynolds numbers, particularly near the critical layers where the wave speed matches the base flow speed.

\subsection{Historical Evolution of Numerical and Asymptotic Solvers}
Before widespread use of electronic computers, solving the Orr-Sommerfeld equation was a premier challenge in applied mathematics. Early analytical efforts in the 1920s to 1940s relied on asymptotic expansions to resolve the critical layer, pioneered by Heisenberg \citep{heisenberg1924}, Tollmien \citep{tollmien1929}, Schlichting \citep{schlichting1933} and Lin \citep{lin1945}. Detailed descriptions may be found in the texts by Lin \citep{lin1955} and Drazin \& Reid \citep{drazin1981}.

The computer era ushered in new methods for estimating the critical values, but early methods, such as standard shooting or finite difference methods, failed due to the rapidly decaying and growing exponential solutions. In 1953, Thomas \citep{thomas1953} used one of the first digital computers and a specialized finite-difference scheme to calculate the first true numerical threshold, finding the critical Reynolds number, $Re_c \approx 5780$. Low-order finite difference schemes require very large grids for accuracy and suffer from large truncation errors and numerical stiffness issues. In 1971, Orszag \citep{orszag1971} bypassed these difficulties through the introduction of the Chebyshev collocation method to fluid dynamics. By doing so he obtained exponential convergence to a solution, and established the canonical six-digit values (e.g. $Re_c=5772.22$), that have served the fluids community for more than 50 years.

\subsection{Numerical Bottlenecks and Literature Discrepancies}
The numerical investigation of hydrodynamic stability in parallel shear flows relies heavily on the accurate spectral solution of the Orr--Sommerfeld equation. While foundational formulations by Orszag established robust spectral benchmarks, modern efforts continue to push the boundaries of discretization and precision. However, such formulations encounter severe numerical bottlenecks. 

Standard double-precision eigensolvers (e.g., MATLAB's \texttt{eig} or standard Chebyshev collocation schemes) compute the full eigenvalue spectrum in 64-bit arithmetic at $\mathcal{O}(N^3)$ complexity for a single parameter pair. Due to the severe ill-conditioning of the discretized Orr--Sommerfeld matrix, such double-precision implementations suffer from catastrophic round-off accumulation, capping critical parameter resolution at $\approx 10^{-8}$--$10^{-12}$. 

Furthermore, attempts to push stability calculations into quadruple precision are fundamentally bottlenecked when finite-difference Jacobians are used in Newton--Raphson iterations. As target tolerances approach floating-point limits, finite differencing suffers from catastrophic cancellation, introducing numerical errors that destroy quadratic convergence. By replacing these approximations with exact, analytical derivatives, the present formulation bypasses full-spectrum decomposition and eliminates derivative approximation errors entirely. This enables robust quadratic convergence down to 28 digits, approaching the quadruple-precision round-off floor ($\epsilon \approx 1.9 \times 10^{-34}$) and far surpassing the hardware precision limit of standard matrix eigensolvers.

Despite the long history of this benchmark, exact agreement on the critical triplet (critical values for Reynolds number, wavenumber, wavespeed) $(Re_c, \alpha_c, c_r)$ has not been universal in the literature. \citet{takashima1996stability} and \citet{shankar2023} agree with the present result to the full precision they report (6 significant figures for $\alpha_c$ and $c_r$, 8 or 10 respectively for $Re_c$), while \citet{giannakis2009spectral} and \citet{piterskaya2025study} report values of $\alpha_c$ and $c_r$ that diverge starting at the $7^{\text{th}}$ significant figure, despite the latter reporting $Re_c$ to 16 digits. This difficulty in constraining $\alpha$ is consistent with the shallow curvature of $Re(\alpha)$ near the critical point: modest errors in the extremality condition $\partial c_i/\partial \alpha = 0$ propagate weakly into $Re_c$ but directly into $\alpha_c$ and $c_r$.

\subsection{Present Contribution and Scope}
Precise, independently-verified extended-precision values for all three quantities is therefore of continued practical value -- not only for verifying hydrodynamic stability codes, but as a validation target for the hydrodynamic limit of the coupled magnetohydrodynamic (MHD) extensions of the Orr-Sommerfeld problem appearing in the literature, for example \citet{takashima1996stability} and \citet{giannakis2009spectral}.

\begin{table}[htbp]
\centering
\caption{Critical values of the Orr--Sommerfeld eigenproblem for PPF reported in the literature, compared with the present
extended-precision result. Apart from the present results, digits are shown to the precision reported by each source.}
\label{tab:critical_comparison}
\resizebox{\linewidth}{!}{
\begin{tabular}{lllll}
\hline
Source & Year & $Re_c$ & $\alpha_c$ & $c_r$ \\
\hline
Present work & 2026 &
  $5772.221\,816\,209\ldots$ &
  $1.020\,547\,449\ldots$ &
  $0.264\,000\,260\ldots$ \\
\citet{takashima1996stability} & 1996 &
  $5772.2218$ &
  $1.020547$ &
  $0.264000$ \\
\citet{giannakis2009spectral} & 2009 &
  $5772.2218$ &
  $1.020551$ &
  $0.264001$ \\
\citet{shankar2023} & 2023 &
  $5772.221816$ &
  $1.020547$ &
  $0.264000$ \\
\citet{piterskaya2025study} & 2025 &
  $5772.221\,833\,743\,744$ &
  $1.020560$ &
  $0.264001\,725\,002\,158$ \\
\hline
\end{tabular}}
\end{table}

To improve the precision limits, the present work introduces a stringent quad-precision Galerkin framework equipped with analytical Jacobians. By bypassing the 64-bit arithmetic overflow limits and eliminating the condition-number sensitivities inherent in conventional collocation approaches, this formulation optimizes for the critical triple to very high precision. The resulting framework converges to the minimum of the neutral stability curve at $Re_c = 5772.2218162...$, establishing a definitive benchmark for linear shear-flow stability.

Using such strategies, this work presents two independent quad-precision methods -- a global spectral Galerkin scheme and a local Riccati integration scheme \citep{davey1977} -- to update the three critical parameters to 28-digit precision. These both use exact analytic Jacobians or sensitivities where necessary for high precision. Operating on fundamentally distinct mathematical formulations, these two methods provide a robust cross-verification of the results. While this level of precision is far higher than needed for any foreseeable physical or modeling applications, the resulting high-precision critical parameters establish an authoritative benchmark for the fluid mechanics community, and the open-source implementation of both methods provides a general framework for high-precision stability analysis.

\section{Formulation}
\subsection{Governing equations}
This numerical study of PPF considers incompressible fluid flow in a two-dimensional (2D) channel, $-H<\hat y<H$. The system is non-dimensionalised using the fluid density $\rho$, channel half-height $H$, and centerline Poiseuille velocity $U_0$. This defines the Reynolds number $Re=U_0 H \rho/\mu$ for dynamic viscosity $\mu$, and dimensionless channel width $-1<y<1$.

The base flow $\bm{u_0}=U(y) \mathbf{e_x}$ has the steady parabolic velocity profile $U(y)=1-y^2$, driven by a constant pressure gradient $P_0$. To investigate the linear stability of this flow, we consider the addition of infinitesimal velocity and pressure perturbations $\bm{\tilde{u}}=(\tilde{u},\tilde{v})$ and $\tilde{p}$ respectively.

The total velocity and pressure fields are decomposed into their background states, $\bm{u_0}$ and $P_0$, plus perturbations, and represented as $\bm{u}=\bm{u_0}+\bm{\tilde{u}}$ and $p = P_0+\tilde{p}$ respectively. These are substituted into the dimensionless Navier-Stokes equations, and non-linear terms of order $\lVert \bm{\tilde{u}} \rVert^2$ are ignored. This yields the linearised continuity and momentum equations:

\begin{align}
    \label{eq:continuity}
    \nabla \cdot \bm{\tilde{u}} &= 0,\\
    \frac{\partial \bm{\tilde{u}}}{\partial t}+(\bm{u_0}\cdot\nabla)\bm{\tilde{u}}+(\bm{\tilde{u}}\cdot\nabla)\bm{u_0}&=-\nabla \tilde{p}+\frac{1}{Re}\nabla^2\bm{\tilde{u}}.
    \label{eq:NS1}
\end{align}
The incompressibility constraint (\ref{eq:continuity}) is implicitly satisfied by introducing a perturbation streamfunction $\psi(x,y,t)$, where $\tilde{u}=\partial\psi/\partial y$ and $\tilde{v}=-\partial\psi/\partial x$. Taking the curl of the momentum equation to eliminate the pressure term leads to the linearized vorticity equation:
\begin{align}
\left(\frac{\partial}{\partial t}+U\frac{\partial}{\partial x}\right)\nabla^2\psi-U''\frac{\partial\psi}{\partial x}=\frac{1}{Re}\nabla^4\psi.
 \label{eq:vort1}
\end{align}
Assuming normal mode solutions of the form $\psi(x,y,t)=\phi(y)e^{i\alpha(x-ct)}$, where $\alpha$ is the real streamwise wavenumber, $c$ is the complex wave speed, and $\phi(y)$ is the complex amplitude, we arrive at the classical Orr-Sommerfeld equation:
\begin{align}
(U-c)(\phi''-\alpha^2\phi)-U''\phi=\frac{1}{i\alpha Re}(\phi''''-2\alpha^2\phi''+\alpha^4\phi), \label{eq:OS}
\end{align}
subject to the no-penetration and no-slip boundary conditions at the channel walls:$$\phi(\pm 1)=\phi'(\pm 1)=0.$$

\subsection{Code Development and Availability}
\label{sec:code_dev}
The numerical solver scripts used in this work were developed by the author with structural code-generation and optimization assistance from Claude Sonnet 5. The Large-Language-Model (LLM) assisted suggestions were restricted to algorithmic structure and snippet generation; all code was subsequently audited, extensively modified, and vetted via unit testing by the author to ensure physical and mathematical accuracy. The final verified codebase is openly accessible. The two methods used for this study are publicly available on GitHub at
\newline
 \href{https://github.com/StephenJWalters/ppf\_critical\_point\_finder\_galerkin}{https://github.com/StephenJWalters/ppf\_critical\_point\_finder\_galerkin}.
\newline
 \href{https://github.com/StephenJWalters/ppf_critical\_point\_finder\_riccati}{https://github.com/StephenJWalters/ppf\_critical\_point\_finder\_riccati}.

\section{Method 1: Spectral Galerkin method and Inverse Iteration}
To solve the Orr-Sommerfeld equation, we expand the complex wall-normal amplitude function $\phi(y)$ as a truncated series of basis functions:$$\phi(y) = \sum_{n=1}^{N_{modes}} a_n G_n(y),$$ where $a_n$ are the complex spectral coefficients. To ensure strict adherence to the physics of the domain, the boundary conditions are enforced by construction. We utilize the `No-Slip-Chebyshev' (NSC) basis functions, defined as:$$G_n(y) = \frac{T_{n-1}(y)-T_{n+1}(y)}{n} + \frac{T_{n+3}(y)-T_{n+1}(y)}{n+2},$$ where $T_n(y) = \cos(n \arccos y)$ represents the standard Chebyshev polynomials of the first kind. As detailed by Walters \& Forbes \citep{waltersforbes2026}, these specific linear combinations of Chebyshev polynomials inherently satisfy the required no-slip wall conditions (as well as the no-penetration condition), $G_n(\pm 1) = G_n'(\pm 1) = 0$, eliminating the need for boundary condition patching or tau-methods.The continuous domain is discretized onto a dense Chebyshev-Gauss-Lobatto quadrature grid, $y_j = -\cos(\pi (j-1) / (N_y-1))$, where $N_y \geq N_{modes}+3$, to ensure exact integration of the inner products.  In the current work an overly generous margin was used, $N_y=5 N_{modes}$ as the quadrature cost was negligible. Substituting the expansion into the Orr-Sommerfeld equation and applying a Galerkin projection via exact Chebyshev quadrature yields a discrete generalized eigenvalue problem of the form:$$A \bm{v} = c Re B \bm{v},$$ in which $\bm{v}$ is the vector of unknown spectral coefficients, and matrices $A$ and $B$ are assembled directly from the analytical derivatives of $G_n(y)$ evaluated on the quadrature grid.

The discrete operators are constructed by evaluating weighted $L^2$ inner products over the domain $y \in [-1, 1]$ with respect to the Chebyshev weight function $w(y) = 1 / (\pi \sqrt{1 - y^2})$. We define the continuous inner product as:$$\langle f, g \rangle_w = \int_{-1}^{1} w(y) f(y) g(y) \, dy.$$The entries of the mass matrix $I^{(0)}$, differential operators $D^{(2)}$ and $D^{(4)}$, and base-flow interaction matrices $M^{(U)}$ and $M^{(U2)}$ are given by:\begin{align*}I^{(0)}_{m,n} &= \langle G_m, G_n \rangle_w, \\
D^{(2)}_{m,n} &= \langle G_m, G_n'' \rangle_w, \quad &D^{(4)}_{m,n} &= \langle G_m, G_n'''' \rangle_w, \\
M^{(U)}_{m,n} &= \langle G_m, U G_n \rangle_w, \quad &M^{(U2)}_{m,n} &= \langle G_m, U G_n'' \rangle_w.\end{align*}

In the numerical implementation, these continuous integrals are evaluated using quadruple precision via Clenshaw--Curtis quadrature on the Chebyshev--Gauss--Lobatto points \citep{ClenshawCurtis1960}, which simplifies the assembly of each matrix to a single matrix multiplication between the basis function arrays and their weighted transposes. In this approach we ignore the sparsity of the discrete operators, along with the associated potential numerical efficiency, and treat all matrices as dense. This allows for a simpler implementation using dense linear algebra, rather than by constructing the operators using their analytic recurrence relations. A higher efficiency implementation along those lines is left as a potential future direction, the focus here being precision and convergence.

Because $U''(y) = -2$ for plane Poiseuille flow, the term in (\ref{eq:OS}) involving the second derivative of the base flow simplifies to a scalar multiple of the mass matrix. Using these operators, the continuous Orr-Sommerfeld equation is mapped directly into the generalized algebraic eigenvalue problem $A \bm{v} = c Re B \bm{v}$. The discrete system matrices $A$ and $B$ are assembled in exact closed form:$$A = D^{(4)} - 2\alpha^2 D^{(2)} + \alpha^4 I^{(0)} - i\alpha Re \left( M^{(U2)} - \alpha^2 M^{(U)} + 2I^{(0)} \right),$$ $$B = -i\alpha \left( D^{(2)} - \alpha^2 I^{(0)} \right).$$A key advantage of this Galerkin framework is that it permits the exact analytical derivation of the system Jacobians. The first- and second-order derivatives of the matrices with respect to the continuous parameters $\alpha$ and $Re$ (e.g., $\partial A/\partial \alpha$, $\partial A/\partial Re$, $\partial^2 A/\partial \alpha^2$) are evaluated as direct algebraic combinations of the static, pre-computed projection matrices. By circumventing numerical finite-difference approximations entirely at the matrix assembly level, the system ensures that the eigenvalue problem and its sensitivities are resolved to the round-off floor of 128-bit arithmetic.

\subsection{Quad-Precision Shift-Invert Inverse Iteration}
Standard linear algebra packages (e.g., LAPACK) do not natively support 128-bit floating-point arithmetic. While there are specialised packages with multiple-precision support, the task here was confined to finding a small number of eigenvalues near a known shift. Therefore, a custom eigensolver was developed to resolve the generalized eigenvalue problem $A \bm{v} = c Re B \bm{v}$ while preserving the quadruple precision of the assembled matrices.

The target eigenvalue $c$ and its corresponding eigenvector $\bm{v}$ are isolated using a shift-invert inverse iteration algorithm. For a given shift $\sigma$, the iteration updates the eigenvector approximation via:$$(A - \sigma Re B) \bm{v}_{k+1} = Re B \bm{v}_k.$$

To solve this linear system at each iteration step, we implemented a custom complex LU factorization with partial pivoting, operating strictly in 128-bit floating-point (quadruple) precision. Because the target wave speed $c$ changes smoothly during the search for the critical point, the shift $\sigma$ is continuously warm-started using the converged eigenvalue from the previous Newton iteration. This dynamic shifting strategy yields highly rapid convergence, typically requiring only 3 to 5 inverse iteration steps per system solve. Once the eigenvector $\bm{v}$ converges, the corresponding complex eigenvalue $c$ is extracted using the generalized Rayleigh quotient: $$c = \frac{\bm{v}^* A \bm{v}}{Re (\bm{v}^* B \bm{v})},$$ where $\bm{v}^*$ denotes the conjugate transpose.

To robustly measure the convergence of the inverse iteration, we employ a phase-invariant metric. Because the complex phase of the eigenvector is arbitrary, directly comparing $\bm{v}_{k+1}$ to $\bm{v}_k$ is error-prone. Instead, convergence is declared when the norm of the component of $\bm{v}_{k+1}$ that is orthogonal to $\bm{v}_k$ falls below a strictly defined quadruple-precision tolerance threshold, $10^{-23}$, ensuring the eigenvector is resolved to the chosen residual tolerance of $4 \times 10^{-29}$. The tolerance value $10^{-23}$ was determined empirically in the course of the grid-refinement study -- larger values gave fewer correct digits, while smaller values could not be satisfied, as the residual was already limited by the quadruple-precision round-off floor.

\subsection{Critical Point Search via Newton-Raphson Iteration}
To locate the critical point where the flow transitions to instability, we solve a two-dimensional root-finding problem defined by the conditions of neutral stability and extremality, $F(Re, \alpha) = (c_i, \partial c_i / \partial \alpha)^T = (0, 0)^T$, using Newton-Raphson iteration.

By differentiating the generalized eigenvalue problem $A \bm{v} = c Re B \bm{v}$ with respect to the parameters $Re$ and $\alpha$, we obtain the derivatives of the eigenvalue $c$ directly from the adjoint relationship. Specifically, the first-order derivative is given by:$$\frac{\partial c}{\partial \lambda} = \frac{\bm{w}^* \left( \frac{\partial A}{\partial \lambda} - c Re \frac{\partial B}{\partial \lambda} - c B \frac{\partial Re}{\partial \lambda} \right) \bm{v}}{\bm{w}^* (Re B) \bm{v}},$$
where $\bm{w}$ is the adjoint eigenvector, and $\lambda \in \{Re, \alpha\}$.

Because the matrices $A$ and $B$ and their derivatives $\partial A/\partial \alpha, \partial^2 A/\partial \alpha^2$, etc., are available in exact closed-form algebraic expressions, we are able to compute the Jacobian analytically, avoiding truncation errors associated with finite difference approximations. This use of analytical derivatives in the Newton iteration ensures quadratic convergence to the critical point, with final residuals reaching $\mathcal{O}(10^{-32})$, effectively resolving the critical parameters to the limits of 128-bit floating-point precision.

\section{Method 2: Riccati Transformation and Sensitivity Integration}
To provide a robust secondary validation, the stability problem is solved using a matrix Riccati transformation method. This approach circumvents the numerical stiffness of the Orr--Sommerfeld equation by transforming the fourth-order boundary-value problem into an initial-value problem for a nonlinear matrix Riccati variable \citep{davey1977}.

We may write the Orr--Sommerfeld equation (\ref{eq:OS}) as:
$$\phi'''' - e(y)\phi'' - d(y)\phi = 0,$$
where, noting that $U''=-2$, the complex coefficients $d(y)$ and $e(y)$ are given by:
$$e(y) = 2\alpha^2 + i\alpha Re(U - c), \quad d(y) = -\alpha^4 - i\alpha Re \left[ \alpha^2(U - c) - 2 \right].$$

We then define a $2 \times 2$ complex Riccati matrix variable $R(y)$, whose evolution across $y \in [-1, 1]$ is governed by:
$$\begin{aligned}
R_{11}' &= R_{21} - R_{12}(d R_{11} + e), \\
R_{12}' &= R_{22} - R_{11} - d R_{12}^2, \\
R_{21}' &= 1 - R_{22}(d R_{11} + e), \\
R_{22}' &= -R_{21} - d R_{12} R_{22}.
\end{aligned}$$

The lower wall boundary conditions $\phi(-1) = \phi'(-1) = 0$ are identically enforced by initializing $R(-1) = \mathbf{0}_{2 \times 2}$. Integrating to the upper wall $y = 1$, the condition for a non-trivial solution satisfying $\phi(1) = \phi'(1) = 0$ yields the scalar eigenvalue condition $F(c) = \det(R(1)) = R_{11}R_{22} - R_{12}R_{21} = 0$.

Parameter sensitivity matrices $S^{\lambda} = \partial R / \partial \lambda$ for $\lambda \in \{Re, \alpha, c\}$ are integrated concurrently in a single pass alongside $R(y)$, from initial conditions $S^{\lambda}(-1) = \mathbf{0}_{2 \times 2}$. Differentiating $\det(R)$ yields the exact derivative of the boundary condition:
$$\frac{\partial F}{\partial \lambda} = R_{22} S^\lambda_{11} - R_{21} S^\lambda_{12} - R_{12} S^\lambda_{21} + R_{11} S^\lambda_{22}.$$

The coupled system for $R$ and its sensitivities $(S^{Re}, S^{\alpha}, S^c)$ is integrated numerically using explicit high-order Runge--Kutta (RK) schemes (such as Zhang's 16-stage, $10^{\text{th}}$-order explicit RK scheme \citep{zhang2019} and Feagin's $14^{\text{th}}$-order 35-stage scheme \citep{feagin2012}) implemented in quadruple precision. At this precision level, the highest order scheme was significantly faster.

The critical point is determined via a two-level nested Newton iteration:
\begin{enumerate}
    \item \textbf{Inner Loop:} For fixed $(Re, \alpha)$, the eigenvalue $c$ is determined using a 1D Newton iteration on $F(c) = 0$. Using the sensitivity $S^c(1)$ (derived analytically and integrated numerically), the derivative $\partial F / \partial c$ is computed without finite-difference approximations, yielding quadratic convergence to $c$.
    \item \textbf{Outer Loop:} The conditions $f_1 = c_i = 0$ and $f_2 = \partial c_i / \partial \alpha = 0$ are solved via a 2D Newton update on $(Re, \alpha)$. The first derivatives $\partial c_i / \partial Re = \text{Im}(-\frac{\partial F / \partial Re}{\partial F / \partial c})$ and $f_2 = \text{Im}(-\frac{\partial F / \partial \alpha}{\partial F / \partial c})$ are obtained analytically without finite differences. The second-order curvature terms ($\partial^2 c_i / \partial \alpha^2$ and $\partial^2 c_i / \partial \alpha \partial Re$) are evaluated via central finite differences of these analytic first derivatives across a small step $h_\alpha = 10^{-9}$.
\end{enumerate}
By eliminating numerical differentiation in the inner loop and first-order parameter sensitivities, this formulation resolves critical parameters to near 128-bit floating-point precision.

\section{Results: Determination of Critical Stability}
To verify the convergence of the critical point $(Re_c, \alpha_c)$ to the limits of quadruple precision, we perform a dual-method analysis. We compare the results of the NSC spectral Galerkin approach (Method 1) against the analytic Riccati integration (Method 2).

\begin{figure}
\centering
\includegraphics[width=\textwidth]{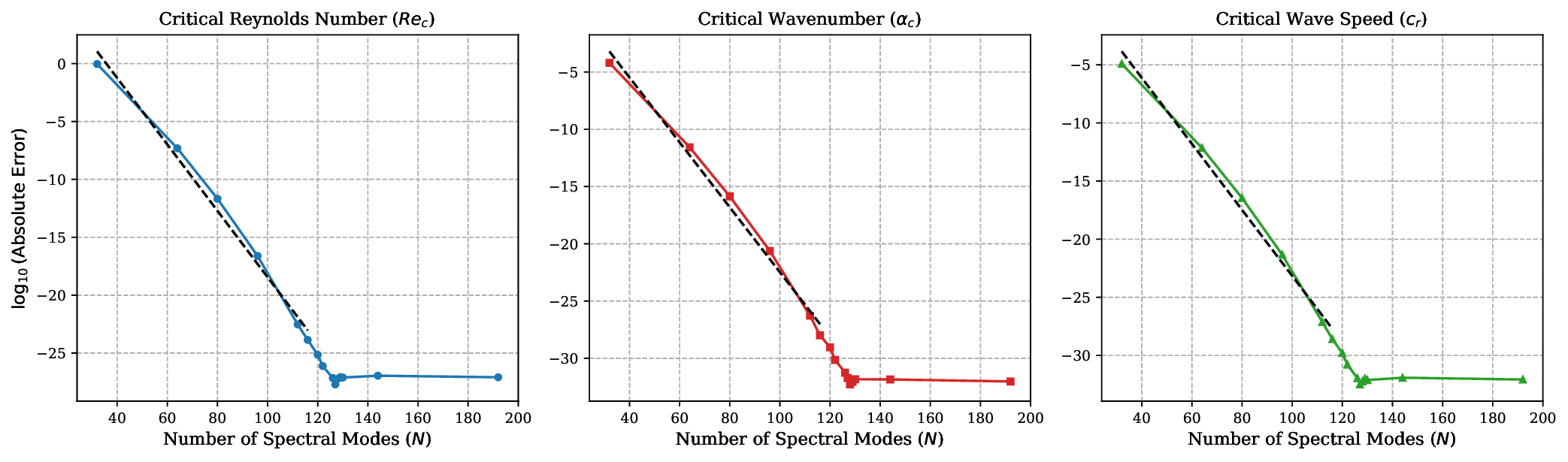}
\caption{Spectral convergence of critical point values with number of modes $N_{modes}$, using the spectral Galerkin method.}
\label{Figure1}
\end{figure}

The spectral convergence of the Chebyshev Galerkin formulation is illustrated in Figure \ref{Figure1} via the decay of absolute errors relative to the 28-digit reference solution. On a semi-logarithmic scale, the error curves exhibit a distinct concave-down curvature prior to reaching the quad-precision floor near $N \approx 128$. This super-exponential convergence rate -- demonstrating error decay outperforming standard geometric bounds -- reflects the absence of any algebraic error component contaminating the exponential convergence, and highlights the high smoothness of the Chebyshev expansion coefficients across the critical layers. Beyond $N = 128$, the residuals flatten uniformly after 30 digits, matching the precision limit of the 128-bit quadruple-precision floating-point environment and confirming that discretization errors have been driven well below the round-off floor.

For Method 1, the results for the critical Reynolds number are also shown as a convergence study in Table \ref{tab:spectralconvergence}. It is seen that 128 spectral coefficients is enough, and larger values of $N_{modes}$ do not provide further precision. The value found has converged to 30 significant figures, out of a possible 33, the remaining three digits holding accumulated roundoff errors from the many floating point operations. The increase in number of correct digits as $N_{modes}$ goes from 32 to 64 to 128 shows exponential convergence, as expected from a spectral method. Runtime for 128 modes was 4.3 seconds on an Intel Core i7-7700K processor (4.2 GHz).

\begin{table}[h]
\centering
\begin{tabular}{cl}
\hline
$N_{modes}$ & $Re_c$ \\
\hline
  256 & $\mathbf{5772.22181620969823713375856640255}$ \\
\hline
   32 & $     577\ldots$ \\
   64 & $     5772.221816\ldots$ \\
  128 & $     5772.22181620969823713375856640\ldots$ \\
  144 & $     5772.22181620969823713375856640\ldots$ \\
  192 & $     5772.22181620969823713375856640\ldots$ \\
\hline
\end{tabular}
\caption{Spectral convergence of critical point values with number of modes $N_{modes}$, using the spectral Galerkin method.}
\label{tab:spectralconvergence}
\end{table}

Converged values for the other two critical values are 
\newline
$\alpha_c=1.0205\, 47449 \,28533 \,35062 \,49464 \,00664 \,13$,
\newline
$c_r=0.26400\, 02604 \,78798 \,44333 \,93402 \,65061 \,7$.

\begin{figure}
\centering
\includegraphics[width=\textwidth]{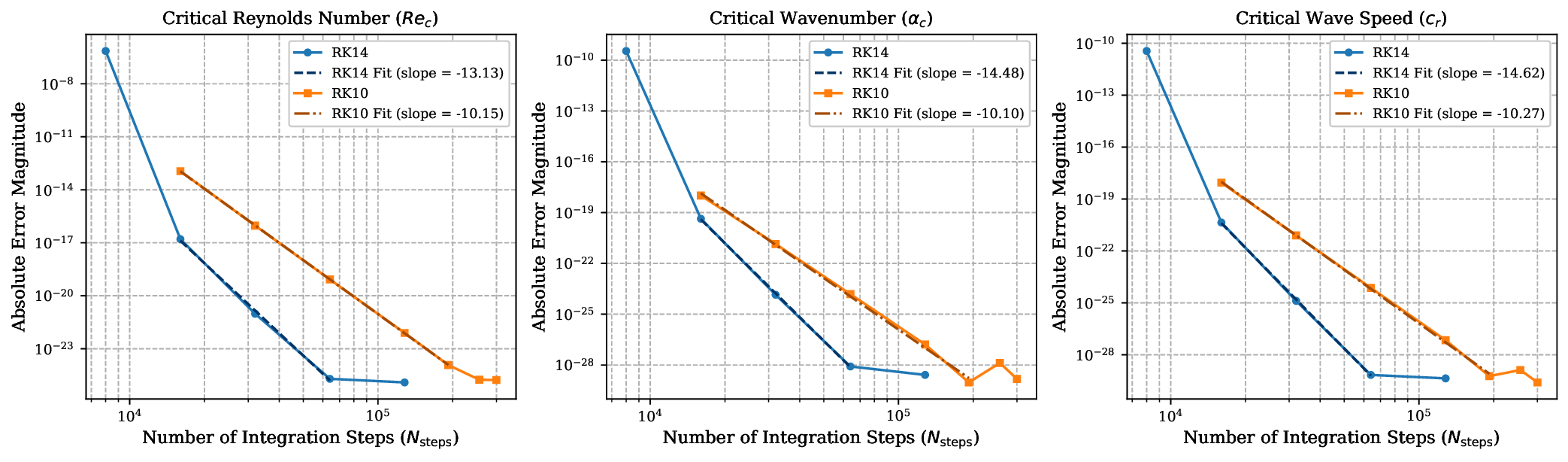}
\caption{Algebraic convergence of critical point values with number of integration steps $N_{steps}$, using a Runge-Kutta methods of order 10 and 14.}
\label{Figure2}
\end{figure}

We look to validate these results with an independent method. Employing Method 2, we vary the spatial resolution until the results have converged. The convergence is shown in Figure \ref{Figure2}, where the fitted lines have the expected slopes of approximately -10 and -14 respectively, for the $10^{\text{th}}$-order and $14^{\text{th}}$-order methods.

Table \ref{tab:riccaticonvergence} shows that maximum precision for this method is found with a spatial grid of 256000 points or more, for the $10^{\text{th}}$-order scheme (16 stages), or 64000 spatial steps for the $14^{\text{th}}$-order method (35 stages). The higher-order scheme has thus reduced runtime (proportional to stages $\times$ steps) by a factor of two, taking 8270 seconds. We note that for the level of precision required, the highest-order method available (14) was fastest for Method 2, but still took 2000 times as long as the spectral Galerkin method.

\begin{table}[h]
\centering
\begin{tabular}{cl}
\hline
$N_y$ & $Re_c$ \\
\hline
  256 mode reference & $\mathbf{     5772.22181620969823713375856640255}$ \\
\hline
$\mathbf{method}$ & $\mathbf{10^{th}-order}$ \\ 
\hline
  16000 & $     5772.221816209698\ldots$ \\
  32000 & $     5772.221816209698237\ldots$ \\
  64000  & $     5772.221816209698237133\ldots$ \\
  128000 & $     5772.221816209698237133758\ldots$ \\
  256000 & $     5772.221816209698237133758566\ldots$ \\
  512000 & $     5772.221816209698237133758566\ldots$ \\
  1024000 & $     5772.221816209698237133758566 \ldots$ \\
  \hline
$\mathbf{method}$ & $\mathbf{14^{th}-order}$ \\
\hline 
   8000 & $     5772.221816\ldots$ \\
   16000 & $     5772.2218162096982371\ldots$ \\
   32000 & $     5772.22181620969823713375\ldots$ \\
   64000 & $     5772.221816209698237133758566\ldots$ \\
  128000 & $     5772.221816209698237133758566\ldots$ \\
\hline
\end{tabular}
\caption{Convergence of critical point values with number of integration steps $N_y$, using the Riccati shooting method with two different high-order RK methods.}
\label{tab:riccaticonvergence}
\end{table}

Both frameworks yield identical values to 28 significant figures, providing robust numerical evidence for the precision of these results. The same pattern is seen for the corresponding critical values of $\alpha_c$ and wavespeed $c_r$. We may therefore be confident that we have found the true critical values to at least 28 significant figures. These values are:

\begin{equation*}
\begin{aligned}
Re_c    &= 5772.2\,21816\,20969\,82371\,33758\,566 \\
\alpha_c &= 1.0205\,47449\,28533\,35062\,49464\,006 \\
c_r     &= 0.26400\, 02604 \,78798 \,44333 \,93402 \,651
\end{aligned}
\end{equation*}

Note that these values agree with the commonly quoted six-significant-figure Orszag values \citep{orszag1971}, apart from $\alpha_c$. Our value, $\alpha_c=1.020547449…$, differs from Orszag's $1.02056 \pm 0.00001$ in the sixth digit, lying just outside his quoted uncertainty. This new value agrees with that of Shankar \& Shivakumara \citep{shankar2023}, their Table 1, $\Lambda=0$, and of \citet{takashima1996stability} to the level of precision quoted.

\section{Conclusion}
\label{sec:conclude}

\subsection{Dual-Framework Numerical Methodology}
Two quad-precision methods have been developed and cross-validated, such that each independently computes the critical stability parameters for PPF. The first method uses a global No-Slip Chebyshev basis within a spectral Galerkin framework. This is used to solve the discretized  generalized eigenvalue problem directly using 2D Newton iteration. The second method is a local, analytically sensitive Riccati matrix integration scheme, marched in space using a $10^{\text{th}}$- or $14^{\text{th}}$-order RK scheme.

While using completely different approaches, these two methods agree to 28 significant figures on the values of the critical triple $Re_c,\alpha_c,c_r$. In both methods, the numerics have been pushed to convergence by increasing spectral modes or reducing grid-spacing respectively.

\subsection{Algorithmic Efficiency and Analytic Jacobians}
During the investigation, it became evident that the use of analytic Jacobians and residuals was essential in pushing out the number of correct digits (by direct differentiation of the matrix operators in Method 1, or by propagating them alongside the Riccati integration in Method 2). In an early version of the code, it was found that use of eighth-order finite differencing in the Jacobian was insufficient for the 28-digit precision. Additionally the analytic approach was significantly faster, due to far fewer function calls (even a second-order stencil requires three calls). Avoiding finite-differencing for the parameter sensitivities allowed the Newton iteration to achieve true quadratic convergence without stalling at a much larger round-off limit.

\subsection{Extended-Precision Benchmark and Software Verification}
Careful numerics and the use of quad-precision arithmetic throughout have allowed the extension of Orszag's canonical six-digit benchmarks to 28 digits, including the correction of Orszag's sixth digit in $\alpha_c$. This correction was also found by \citet{takashima1996stability} and by \citet{shankar2023}, who give the critical triple to six decimal places. We agree with their results, but have greatly extended the precision.

This high precision result serves as definitive evidence that the analytical Jacobian formulation and Newton residual evaluations are correct, demonstrating that the severe ill-conditioning famously associated with the Orr--Sommerfeld problem can be completely controlled in a robust, verifiable framework.

The 28-digit values found for the critical triple may now serve as the new definitive benchmark for the stability of Plane Poiseuille Flow. Additionally, the dual-framework structure is presented as an excellent stress test for verifying the accuracy, boundary condition implementation, and spectral resolution limits of future hydrodynamic stability codes. While demonstrated here for Plane Poiseuille flow, the analytical-sensitivity formulation developed in both the spectral and Riccati frameworks is directly transferable to other shear flows; extending this quadruple-precision benchmark suite to semi-infinite boundary layers (such as Blasius flow) represents a natural direction for future work.

Beyond methodological validation, the high-precision reference values presented here provide practical utility for algorithm development. Specifically, establishing benchmark parameters to 28 decimal places allows researchers testing double-precision ($10^{-16}$) or extended-precision solvers to plot true convergence errors down to machine epsilon without contamination from benchmark truncation. This enables sensitive detection of internal code consistency, convergence rates, and subtle round-off errors at much lower mode counts than was previously possible with older six-digit references.

The resulting triplet provides an unambiguous, independently-verified reference against which future stability solvers -- including MHD solvers tested in their pure hydrodynamic limit -- can be validated to high precision.

\section*{Acknowledgments}
I am grateful to my friend and colleague, Professor Larry Forbes for helpful comments on the initial draft of this paper.

\section*{Statement on the use of large language models}
During the preparation of this work the author used Google Gemini 3.1 Pro and Claude Sonnet 5 to suggest improvements to sentence-level phrasing and formatting in author-drafted technical descriptions of the numerical methodology, and for code-generation as described in \ref{sec:code_dev}. All LLM-suggested wording was critically reviewed, revised, and verified against the underlying methodology by the author, who takes full responsibility for the accuracy and content of the publication.

\bibliographystyle{plainnat}

\end{document}